\documentclass[envcountsame,envcountsect,runningheads]{llncs}

\usepackage{silence}
\usepackage[utf8]{inputenc}
\usepackage[
  colorlinks=true,
  linkcolor=blue,
  filecolor=blue,
  urlcolor=blue,
  citecolor=blue,
]{hyperref}
\hypersetup{bookmarksdepth=subsection}
\usepackage{color}

\usepackage[T1]{fontenc}
\usepackage{tikz}
\usetikzlibrary{cd}
\usepackage{cite}
\usepackage{amsfonts}
\usepackage{amsmath}
\usepackage{amssymb}
\usepackage{proof}
\usepackage{cmll}
\usepackage{xspace}
\allowdisplaybreaks

\newcommand\Q{\mathcal{Q}}
\newcommand\super[2]{\langle #1,#2\rangle^\odot}
\newcommand\pair[2]{\langle #1,#2\rangle}
\newcommand\inl{\mathsf{inl}}
\newcommand\inr{\mathsf{inr}}
\newcommand\inlr[2]{\mathsf{inlr}(#1,#2)}
\newcommand\sumwe[2]{\ensuremath{\nabla_{\!\!#1#2}}}

\newcommand\coproducto[2]{{[}#1,#2{]}}
\newcommand\Hom[2]{\mathsf{Hom}({#1},{#2})} 
\newcommand\TheCat{\ensuremath{\mathbf{Mag}_{\mathbf{Set}}}\xspace}
\newcommand\TheCatAlg{\ensuremath{\mathbf{AMag}^{\mathcal{S}}_{\mathbf{Set}}}\xspace}
 \newcommand\cp{\ensuremath{\mathbin{\raisebox{0.3ex}{\scalebox{0.8}{\ooalign{$\bigcirc$\cr\hfil${\scalebox{0.82}{$\dashv\vdash$}}$}}}}}}
\begin{document}
\title{Towards a Computational Quantum Logic}
\subtitle{An Overview of an Ongoing Research Program}
%
%
\author{Alejandro Díaz-Caro\inst{1,2}}
\authorrunning{A. Díaz-Caro}
%
\institute{
  Université de Lorraine, CNRS, Inria, LORIA, F-54000 Nancy, France
\and
  Universidad Nacional de Quilmes, Bernal, Buenos Aires, Argentina
  \\
  \email{alejandro.diaz-caro@inria.fr}
}
\maketitle              
\begin{abstract}
This invited paper presents an overview of an ongoing research program aimed at
extending the Curry-Howard-Lambek correspondence to quantum computation. We
explore two key frameworks that provide both logical and computational
foundations for quantum programming languages.

The first framework, the Lambda-$S$ calculus, extends the lambda calculus by
incorporating quantum superposition, enforcing linearity, and ensuring
unitarity, to model quantum control. Its categorical semantics establishes a
structured connection between classical and quantum computation through an
adjunction between Cartesian closed categiries and additive symmetric monoidal
closed categories.

The second framework, the $\mathcal{L}^{\mathbb C}$ calculus, introduces a proof language
for intuitionistic linear logic augmented with sum and scalar operations. This
enables the formal encoding of quantum superpositions and measurements, leading
to a computational model grounded in categorical structures with biproducts.

These approaches suggest a fundamental duality between quantum computation and
linear logic, highlighting structural correspondences between logical proofs
and quantum programs.  We discuss ongoing developments, including extensions to
polymorphism, categorical and realizability models, as well as the integration
of the modality ${!}$, which further solidify the connection between logic and
quantum programming languages.

\keywords{Quantum Computing \and Lambda calculus \and Linear Logic \and Categorical Semantics.}
\end{abstract}

\section{Motivation}
The term \emph{quantum logic} was first introduced by Garrett Birkhoff and John
von Neumann nearly a century ago~\cite{BirkhoffVonNeumann36} to describe a
formal system inspired by the structure of quantum theory.  Unlike classical
logic, which is based on Boolean algebra, quantum logic replaces the
distributive law with a weaker condition, leading to an orthocomplemented
lattice structure, that is, a bounded lattice in which each element has a
unique orthocomplement satisfying involution, order-reversing, and complement
laws.  This formulation aligns with the mathematical properties of quantum
mechanics, where propositions correspond to projections in a Hilbert space.

While quantum logic has been explored as a foundational framework for reasoning
about quantum mechanics, its connection to computation remains underdeveloped.
In contrast, the interplay between intuitionistic logic, typed lambda calculus,
and Cartesian closed categories has been extensively studied, with the
Curry-Howard-Lambek correspondence~\cite{Howard1980,Lambek1980,SorensenUrzyczyn06,Crole93} revealing
deep structural connections among these domains.

To extend this correspondence to quantum computation, a logical foundation that
captures its structure is needed. This paper aims to provide an overview of a
long-term research program focused on developing a \emph{computational} quantum
logic as a foundation for quantum programming languages. The approach is based
on two key frameworks: the Lambda-$S$
calculus~\cite{DiazcaroDowekRinaldiBIO19,DiazcaroGuillermoMiquelValironLICS19},
an extension of the lambda calculus for quantum computing, and the $\mathcal
L^\mathcal S$ calculus~\cite{DiazcaroDowekMSCS24}, a proof language for
intuitionistic linear logic whose proof terms enable the construction of
quantum programs. The strategy follows a two-step process: first, deriving
logic from computation (via the Lambda-$S$ calculus), and then deriving
computation from logic (via the $\mathcal{L^S}$ calculus), ultimately aiming to converge in the middle.

\section{The \texorpdfstring{Lambda-$S$}{Lambda-S} Calculus: From Computing to Logic}
\subsection{The \texorpdfstring{\emph{Lineal}}{Lineal} Origin}
Lambda-$S$~\cite{DiazcaroDowekRinaldiBIO19} is a quantum lambda
calculus designed to handle quantum superposition and control while preserving
key computational properties such as strong normalization and subject
reduction. This calculus extend the lambda calculus by incorporating algebraic
linearity and type-based constraints to enforce quantum mechanical principles,
particularly the no-cloning theorem.

This approach originates from the algebraic lambda calculus
\emph{Lineal}~\cite{ArrighiDowekLMCS17}. The primary goal of Lineal was to
extend the untyped lambda calculus to effectively express quantum programs,
following a simple yet powerful idea: in the lambda calculus, programs and data
are indistinguishable. Thus, if quantum computing allows for superpositions of
data, the lambda calculus must reflect this by allowing superpositions of
programs. This idea was formalized in Lineal, a lambda calculus where program
superposition is a first-class citizen. Specifically, if $t$ and $r$ are lambda
terms, then so is $\alpha \cdot t + \beta \cdot r$, with $\alpha, \beta \in
\mathbb{C}$.
For example, if $t$ and $r$ are the encodings of $\mathsf{true}$ and
$\mathsf{false}$, respectively, then $\alpha\cdot t+\beta\cdot r$ is the
encoding of the vector $\begin{pmatrix} \alpha\\\beta \end{pmatrix}$, which
represents a superposition of the two values.

The name Lineal derives from the Spanish word for linear, as its reduction
strategy enforces linearity through the following rule:
\begin{equation}
\label{eq:linearity}
(\lambda x.t)(\alpha \cdot v + \beta \cdot w) \longrightarrow \alpha \cdot (\lambda x.t)v + \beta \cdot (\lambda x.t)w
\end{equation}
where $t$ is any term, and $v$ and $w$ are \emph{basis terms}, meaning they are standard values, so
they are not linear combinations of other terms.  This reduction strategy,
known as call-by-base~\cite{AssafDiazcaroPerdrixTassonValironLMCS14}, ensures
that lambda terms such as $\lambda x.t$ are not explicitly required to be
linear but are instead forced to behave linearly through the reduction process.
Since quantum gates are unitary transformations, and thus are inherently
linear, Lineal naturally encodes \emph{measurement-free} quantum programs.

Indeed, the quantum measurement is not linear.  To illustrate this, consider an
orthonormal basis $\mathcal{B} = \{\vec v, \vec w\}$ of $\mathbb{C}^2$ and a
superposition $\alpha\vec{v} + \beta\vec{w}$ with $|\alpha|^2 + |\beta|^2 = 1$.
Measuring this state with respect to $\mathcal{B}$ collapses the state to
$\vec{v}$ with probability $|\alpha|^2$ and to $\vec{w}$ with probability
$|\beta|^2$. This \emph{probabilistic collapse} is inherently non-linear, and
since Lineal enforces linearity in its reduction strategy, it cannot directly
encode quantum measurements.

This limitation motivated the development of the Lambda-$S$ calculus. The key
observation was that if a term $\lambda x.t$ includes a measurement of its
argument, it must consume the entire superposition passed as an argument rather
than distributing it according to rule~\eqref{eq:linearity}.
Lambda-$S$ addresses this by enforcing linearity in the reduction strategy while
allowing terms to consume full superpositions in cases where linearity is
preserved through other means. Inspired by Intuitionistic Linear Logic,
Lambda-$S$ ensures that if $\lambda x.t$ consumes its whole argument, it uses it
exactly once. This is achieved by combining the enforced linearity of the
reduction strategy with the linearity constraints of the type
system~\cite{DiazcaroDowekRinaldiBIO19}.

To accomplish this, Lambda-$S$ introduces a type constructor S(A), where a simple
type A represents a set of basis terms and S(A) denotes its span. This leads to
two distinct $\beta$-reductions based on the type of the variable:
\begin{itemize}
  \item If $x$ has type S$(A)$, then $(\lambda x^{S(A)}.t)r$ reduces in a
    call-by-name fashion, with the type system ensuring that $t$ uses $x$ only
    once.
  \item If $A \neq S(A^{\prime})$, then $(\lambda x^A.t)r$ reduces in
    call-by-base mode.
\end{itemize}
Since the calculus is Church-style, reduction dependent on typing amounts to considering two abstractions.

A later refinement, Lambda-$S_1$~\cite{DiazcaroMalherbeLMCS22}, further enforces norm-preserving constraints on superpositions and guarantees unitarity. This refinement was developed building upon a realizability model proposed earlier~\cite{DiazcaroGuillermoMiquelValironLICS19}. This realizability model addresses the long-standing issue of preserving quantum unitarity in quantum control lambda calculi.

\subsection{Categorical Semantics}\label{sec:adjunction}
\begin{figure}[t]
  \centering
    \begin{tikzcd}[column sep=3mm,row sep=0mm]
      (\mathcal{S},\times,I)\ar[rr,bend left,"{(F,m)}"] & \bot & (\mathcal{V},\otimes,1)\ar[ll,bend left,"{(G,n)}"]
    \end{tikzcd}
  \caption{The adjunction between the categories $\mathcal{S}$ and $\mathcal{V}$}
  \label{fig:adjunction}
\end{figure}
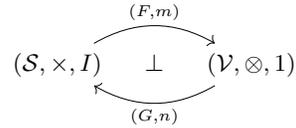

Lambda-$S$ has been given a categorical semantics~\cite{DiazcaroMalherbeACS2020}
through an adjunction between a Cartesian closed category
$(\mathcal{S},\times,I)$ and an additive symmetric monoidal closed category
$(\mathcal{V},\otimes,1)$ (Figure~\ref{fig:adjunction}).

A concrete example of this adjunction is given by the category $\mathbf{Set}$
of sets and functions as $\mathcal{S}$, and the category $\mathbf{Vec}$ of
vector spaces and linear maps as $\mathcal{V}$~\cite{DiazcaroMalherbeMSCS2023}.
In this case, $F$ is the span functor, while $G$ is the forgetful functor.

In this setting, the type constructor $S$ is interpreted by the monad $GF$,
which first transforms a set into a vector space and then forgets its vector
space structure, treating it again as a set.

This implies that, on the one hand, we have tight control over the Cartesian
structure of the model (e.g., duplication), while on the other hand, the world
of superpositions is, in a sense, embedded within the classical
world---determined externally by classical rules until explicitly explored.
This is represented by the following composition of maps:
\[
  GF({A})\times GF({A})\xrightarrow{n} G(F({A})\otimes F({A}))\xrightarrow{G(m)} GF({A}\times {A})
\]
which allows operations within a monoidal structure, explicitly enabling
algebraic manipulation, before returning to the Cartesian product.
Intuitively, using \textbf{Set} and \textbf{Vec} as concrete examples of a
Cartesian and a monoidal category, respectively, this composition corresponds
to taking the Cartesian product of the spans of two sets (seen as sets),
transforming it into their tensor product (in the linear world), and then
returning to the span of the Cartesian product of the original sets.  The
passage through the tensor space allows for the necessary algebraic
manipulation.  This contrasts with linear logic, where the ${!}$ modality
prevents algebraic manipulation; that is, in linear logic, $({!}{A})\otimes
({!}{A})$ remains a product within a monoidal category.

The categorical semantics of Lambda-$S_1$~\cite{DiazcaroMalherbeLMCS22} is
structured around a similar adjunction but diverges from Lambda-$S$ by ensuring
that all terms maintain a unitary norm.

Both calculi serve as foundations for quantum programming languages and
categorical models of quantum computation, bridging classical and quantum
computational paradigms through rigorous mathematical structures. 

\subsection{Extensions and Future Directions}

A recent development in this line of research has been to extend the
superpositions beyond the computational basis---which is essentially what we
assume when choosing a measurement basis of non-superposed values---to include
a diagonal basis~\cite{Monzon25}. In this work, the terms $|0\rangle$ and
$|1\rangle$ are constants in the language, but also $|+\rangle =
\frac{1}{\sqrt{2}}\cdot|0\rangle +\frac{1}{\sqrt{2}}\cdot |1\rangle$ and
$|-\rangle = \frac{1}{\sqrt{2}}\cdot|0\rangle + \frac{-1}{\sqrt 2}\cdot
|1\rangle$. This serves as a proof of concept, demonstrating the ability to
track duplicable basis terms across multiple bases.  

A more general extension of Lambda-$S_1$ to arbitrary bases, following a
realizability model, is currently under development.

The line of research on Lambda-$S$ has demonstrated how to extend the lambda
calculus to quantum computation. In terms of proof theory, it provides a proof
language for intuitionistic logic with a modality $S$, indicating what cannot
be duplicated.

As mentioned earlier, Intuitionistic Linear Logic is
interpreted in a monoidal closed category, with an adjunction to a Cartesian
closed category to account for duplicable data. This is the opposite approach
to that of Lambda-$S$.  

This duality strongly suggests that the structure of quantum computation is the
dual of linear logic and that the Curry-Howard-Lambek correspondence can be
extended to quantum computation---where this dual of linear logic serves as the
logical counterpart, and Lambda-$S$ as the computational counterpart.

\section{The \texorpdfstring{$\mathcal{L^S}$}{L-S} Calculus: From Logic to Computing}
\subsection{The \texorpdfstring{\emph{Sup}}{Sup} Connective}
In the previous section, we introduced Lambda-$S$ by extending the untyped lambda
calculus with modifications necessary to handle quantum computation.
This led to an intuitionistic logic with a novel modality $S$ marking superpositions
and a categorical model structurally related to linear logic.

The next step was to explore the opposite direction:
starting with a well-established logic and modifying it to produce a quantum
programming language within its proof system. Our first concern was to
understand how the non-determinism from the quantum measurement could fit into
a logical framework.

Our starting observation was that a quantum superposition cannot be represented
by a conjunction, as its eliminations are deterministic: one can project either
the first component or the second, but the choice is fixed. Similarly, a
quantum superposition cannot be represented by a disjunction, as its
introductions require only one of the components to be present, and thus, its
elimination can only consider that one.  This led us to introduce a new
connective, which adopts the introduction rules of conjunction and the
elimination rules of disjunction.  This new connective, called sup and denoted
$\odot$, requires both components to be present to be introduced, but its
elimination allows for the non-deterministic choice of one of the components. 
\[
  \infer[\odot_I]{\Gamma\vdash A\odot B}{\Gamma\vdash A &\Gamma\vdash B}
  \qquad
  \qquad
  \infer[\odot_E]{\Gamma\vdash C}{\Gamma\vdash A\odot B &\Gamma,A\vdash C & \Gamma,B\vdash C}
\]
This way, the cut elimination process for the $\odot$ connective is not deterministic, since
\[
  \infer[\odot_E]{\Gamma\vdash C}{\infer[\odot_I]{\Gamma\vdash A\odot B}{\infer{\Gamma\vdash A}{\pi_A} &\infer{\Gamma\vdash B}{\pi_B}} &\infer{\Gamma,A\vdash C}{\pi_C^A} & \infer{\Gamma,B\vdash C}{\pi_C^B}}
\]
can be reduced to either by replacing the proof $\pi_A$ in the proof $\pi_C^A$ or the proof $\pi_B$ in the proof $\pi_C^B$.

This connective was introduced in~\cite{DiazcaroDowekTCS23a},
where we explored its properties and its application to encode quantum
superpositions and measurements.
While bits are usually encoded by $\top\vee\top$, in this logic, we encode qubits by $\top\odot\top$, and the non-determinism present in the elimination rule models the probabilistic nature of quantum measurements. 
However, we went further and demonstrated how to encode matrices and vectors using the $\odot$ connective by introducing two additional rules: one for scalars and another for sums of terms.
\[
  \infer[\textrm{Sum}]{\Gamma\vdash A}{\Gamma\vdash A & \Gamma\vdash A}
  \qquad\qquad
  \infer[\textrm{Scalar}]{\Gamma\vdash A}{\Gamma\vdash A}
\]
The Sum rule enables a deterministic cut-elimination for the $\odot$ connective by summing the results of both paths.
The Scalar rule allows us to transform non-deterministic choices into probabilistic ones. While these two rules are trivally admissible in the logic, they may stop cut-elimination from occurring. Indeed, they could appear in the middle of an introduction and an elimination, as in the following example
\[
  \infer[\odot_E]{\Gamma\vdash C}
  {
    \infer[\textrm{Scalar}]{\Gamma\vdash A\odot B}
    {
      \infer[\odot_I]{\Gamma\vdash A\odot B}
      {
	\infer{\Gamma\vdash A}{\pi_A} &\infer{\Gamma\vdash B}{\pi_B}
      }
    }
    &
    \infer{\Gamma,A\vdash C}{\pi_C^A} & \infer{\Gamma,B\vdash C}{\pi_C^B}
  }
\]
Thus, we also introduced all the necessary commuting rules to ensure the introduction property, that is, guaranteeing that the cut-elimination process does not get stuck.
In this case, for example, we introduced a rule commuting the rule $\textrm{Scalar}$ with the rule $\odot_I$ as follows
\begin{equation}
  \label{eq:commute}
  \vcenter{
    \infer[\textrm{Scalar}]{\Gamma\vdash A\odot B}
    {
      \infer[\odot_I]{\Gamma\vdash A\odot B}
      {
	\infer{\Gamma\vdash A}{\pi_A} &\infer{\Gamma\vdash B}{\pi_B}
      }
    }
  }
  \qquad\longrightarrow\qquad
  \vcenter{
    \infer[\odot_I]{\Gamma\vdash A\odot B}
    {
      \infer[\textrm{Scalar}]{\Gamma\vdash A}
      {	\infer{\Gamma\vdash A}{\pi_A} }
      &
      \infer[\textrm{Scalar}]{\Gamma\vdash B}
      {\infer{\Gamma\vdash B}{\pi_B}}
    }
  }
\end{equation}

The proof language is then extended with a new kind of pair as proof terms for the $\odot$ connective, along with sums of proofs and scalar multiplication, much in the way of Lineal. Additionally, in order to represent vectors, we define every scalar $\alpha \in \mathcal{S}$---the set of scalars (usually taken as $\mathbb{C}$ to represent quantum computing)---as a proof of $\top$.
\[
  \infer[\odot_I]{\Gamma\vdash \super tr:A\odot B}{\Gamma\vdash t:A &\Gamma\vdash r:B}
  \quad
  \infer[\odot_E]{\Gamma\vdash\mathsf{match}^\odot(t,x.r,y.s):C}{\Gamma\vdash t:A\odot B &\Gamma,x:A\vdash r:C & \Gamma,y:B\vdash s:C}
\]
\[
  \infer[\textrm{Sum}]{\Gamma\vdash t+r:A}{\Gamma\vdash t:A & \Gamma\vdash r:A}
  \qquad
  \infer[\textrm{Scalar}(\alpha)]{\Gamma\vdash \alpha\cdot t:A}{\Gamma\vdash t:A}
  \qquad
  \infer[\top_I(\alpha)]{\Gamma\vdash\star_\alpha:\top}{}
\]
Thus, the commuting rule of Equation~\eqref{eq:commute} corresponds to the following rewriting rule of the proof-language
\[
  \alpha.\super tr\longrightarrow \super{\alpha\cdot t}{\alpha\cdot r}
\]

Contrary to Lineal, where the linearity is enforced in the reduction strategy,
we did not forced linearity in the proof-terms with an strategy in the
$\odot$-calculus. Instead, we showed how to encode matrices and vectors with
these contructions.  

First, we defined the proposition $\Q^{\otimes n}$ by induction on $n$ as follows: 
$\Q^{\otimes 0} = \top$ and $\Q^{\otimes (n+1)} = \Q^{\otimes n} \odot \Q^{\otimes n}$.
Then, we have a one-to-one correspondance between the closed irreducible proofs of $\Q^{\otimes n}$ and the vectors of ${\mathbb C}^{2^n}$, associating to each closed irreductible proof-term $t$ of $\Q^{\otimes n}$ a vector $\underline t$ of ${\mathbb C}^{2^n}$ as follows:
\begin{itemize}
  \item If $t = \star_\alpha$, we let $\underline{t} = \left(\begin{smallmatrix} \alpha \end{smallmatrix}\right)$.
  \item If $t = \super{t_1}{t_2}$, we let $\underline{t}$ be the vector with two
blocks $\underline{t_1}$ and $\underline{t_2}$:  $\underline{t} =
\left(\begin{smallmatrix}
  \underline{t_1}\\\underline{t_2} \end{smallmatrix}\right)$.
\end{itemize}

In the same way, to each vector $\vec{v}$ of ${\mathbb C}^{2^n}$, we associate a closed
irreducible proof $\overline{\vec{v}}$ of $\Q^{\otimes n}$.
\begin{itemize}
  \item If $\vec{v} = \left(\begin{smallmatrix} \alpha \end{smallmatrix}\right)$, we let $\overline{\vec{v}} = \star_\alpha$.
  \item If $\vec{v} = \left(\begin{smallmatrix} \vec{v}_1\\ \vec{v}_2\end{smallmatrix}\right)$, where $\vec{v}_1$ and $\vec{v}_2$ are the two blocks of size $\frac n2$, 
we let $\overline{\vec{v}} = \super{\overline{\vec{v}_1}}{\overline{\vec{v}_2}}$.
\end{itemize}

The interesting part is that the Sum and Scalar rules act as a sum of vectors and a scalar multiplication:
\begin{theorem}
  [Sum and Scalar Multiplication~{\cite[Props.~5.6 and 5.7]{DiazcaroDowekTCS23a}}]
  If $t$, $t_1$, and $t_2$ are closed proofs of $\Q^{\otimes n}$, and $\alpha\in\mathbb C$,we have
  \[
    \underline{t_1 + t_2} = \underline{t_1} + \underline{t_2}\textrm{,}\qquad\textrm{ and }\qquad 
    \underline{\alpha\cdot t} = \alpha\underline{t}.
    \tag*{$\triangleleft$}
  \]
\end{theorem}

Moreover, we can encode matrices as stated by the following theorem.
\begin{theorem}[Matrices {\cite[Thm.~5.8]{DiazcaroDowekTCS23a}}]
\label{thm:matrices}
Let $M$ be a matrix with $2^m$ columns and $2^n$ lines, then there
exists a closed proof $t$ of $\Q^{\otimes m} \multimap \Q^{\otimes
  n}$ such that, for all vectors $\vec{v} \in {\mathbb C}^{2^m}$,
  $\underline{t\overline{\vec{v}}} = M \vec{v}$.
  \hfill$\triangleleft$
\end{theorem}

\subsection{The Sum and Scalar Rules in Linear Logic: The \texorpdfstring{$\mathcal{L^S}$}{L-S} Calculus}
In~\cite{DiazcaroDowekMSCS24}, we extended the study of the Sup-calculus to the
context of linear logic, introducing the $\mathcal L\odot^\mathcal S$
calculus. Additionally, we studied a fragment without $\odot$ that retains the
Sum and Scalar rules, which we referred to as the $\mathcal{L^S}$ calculus.
In all these calculi, $\mathcal S$ is a set of scalars, usually, a semiring.

This framework integrates the ideas from the previous section into
intuitionistic multiplicative-additive linear logic (IMALL), establishing a 
formal proof language for this logical system.
The inclusion of scalars and sums ensures that proofs in $\mathcal{L^S}$ 
represent linear maps, and we provided a syntactic proof of this property.
In particular, we proved the converse of Theorem~\ref{thm:matrices}, showing that 
every proof-term of $\Q^{\otimes m} \multimap\Q^{\otimes n}$ represents a
linear map from ${\mathbb C}^{2^m}$ to ${\mathbb C}^{2^n}$.
In fact, the statement of the theorem is more general, as it holds for any proposition as domain.
\begin{theorem}[Linearity {\cite[Thm.~4.10]{DiazcaroDowekMSCS24}}]
  \label{thm:linearity}
  Let 
  $\vdash \lambda x.t:A\multimap\Q^{\otimes n}$, $\vdash r:A$, $\vdash s:A$, and $\alpha,\beta\in\mathbb C$. Then
  \[
    (\lambda x.t)(\alpha\cdot r+\beta\cdot s)\longrightarrow^* v
    \qquad\Longleftrightarrow\qquad
    \alpha\cdot(\lambda x.t)r + \beta\cdot(\lambda x.t)s\longrightarrow^* v
    \tag*{$\triangleleft$}
  \]
\end{theorem}
Moreover, this result also extends to any proposition in the codomain using an observational equivalence $\sim$.
\begin{corollary}[Generalized linearity {\cite[Cor.~4.11]{DiazcaroDowekMSCS24}}]
  \label{cor:linearity}
  Let $\vdash \lambda x.t:A\multimap B$, $\vdash r:A$, $\vdash s:A$, and $\alpha,\beta\in\mathbb C$. Then the following observational equivalence holds
  \[
    (\lambda x.t)(\alpha\cdot r+\beta\cdot s)
    \qquad\sim\qquad
    \alpha\cdot(\lambda x.t)r + \beta\cdot(\lambda x.t)s
    \tag*{$\triangleleft$}
  \]
\end{corollary}

\subsection{Categorical Semantics}
\subsubsection{The Linear Case}\label{sec:catlin}
The $\mathcal L\odot^\mathcal S$ calculus has been assigned a categorical
semantics in a recent draft~\cite{DiazcaroMalherbe24}.  The core idea of
introducing a connective that behaves as both a conjunction and a disjunction,
while also incorporating sums and scalars within a linear logic framework,
aligns naturally with an additive symmetric monoidal closed category. Such a
category, equipped with biproducts, provides a suitable setting for
interpreting this conjunction-disjunction connective, as well as the sum and
scalar operations. 

Introducing a probabilistic operator into a linear language is not straightforward. To aid intuition, consider the concrete category $\mathbf{Vec}$ of vector spaces and linear maps over $\mathbb C$.
Our interpretation cannot use the Powerset Monad, as it is usual to express
non-deterministic effects~\cite{MoggiIC91}, since this monad is not valid in
this category. Such approach would consist in using the Cartesian product of
the several non-deterministic paths, and gather them together into a set.  The
problem is that the needed map $A\times A\xrightarrow{\xi}\mathcal PA$
defined by $\xi(a_1,a_2)=\{a_1,a_2\}$ is not linear and so it is not in the
category.

Our approach was instead inspired by the density matrix quantum formalism (see,
for example,~\cite[Section 2.4]{NC}), wherein we consider the linear combination
of results as a representation of a probability distribution.  Let $t$ be a term
reducing with probability $p$ to $t_1$ and a probability of $q$ to $t_2$, with
$p+q=1$.  
We
interpret $t$ as 
\(
 \sumwe pq(t_1,t_2) = \hat p\cdot t_1 + \hat q\cdot t_2
\), where, if $\hat p$ is the mapping that multiplies its argument by $p$, then
$\sumwe pq$ is defined as $\coproducto{\hat p}{\hat q}$ (Figure~\ref{fig:nablapq}).
\begin{figure}[t]
  \centering
  \begin{tikzcd}[column sep=2cm,row sep=1cm]
      {A}\ar[r,"i_1"]\ar[dr,"\hat p"',sloped] & A+A\ar[d,"\sumwe pq"'] & {A} \ar[dl,"\hat q"',sloped]\ar[l,"i_2"']\\
      &  A &
    \end{tikzcd}
  \caption{The map $\sumwe pq$ as $\coproducto{\hat p}{\hat q}$}
  \label{fig:nablapq}
\end{figure}

This approach is closed to that used for PCF$^{\mathcal R}$~\cite{LairdManzonettoMcCuskerPaganiLICS13}. 
However, in an abstract categorical setting, this means that we needed a category
with biproducts, so we can interpret $\sumwe pq$ as $\nabla\circ(\hat p\oplus
\hat q)$, where $\hat p$ and $\hat q$ are appropriated maps $A\to A$.  For those
scalar maps, we considered the category to also be monoidal, so we can count with
the semiring of scalars $\Hom 11$~\cite{KellyLaplaza80}, where $1$ is the tensor
unit. 
Then, we defined a monomorphism $\mathcal S\to \Hom II$ to interpret the scalars,
which ensures that two proof-terms interpreted by the same map are structurally equivalent.

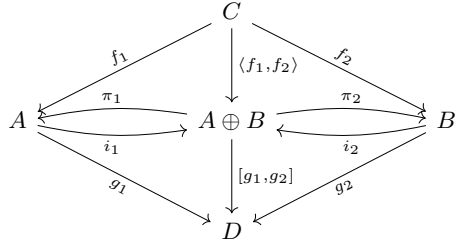
\begin{figure}[t]
  \centering
  \begin{tikzcd}[column sep=2cm,row sep=1cm]
    & C\ar[d,"\pair {f_1}{f_2}"]\ar[dl,"f_1",sloped]\ar[dr,"f_2",sloped] \\
    A\ar[dr,"g_1"',sloped]\ar[r,"i_1"',bend right=10] & A\oplus B\ar[d,"{[g_1,g_2]}"]\ar[r,"\pi_2",bend left=10]\ar[l,"\pi_1"',bend right=10] & B\ar[l,"i_2",bend left=10]\ar[dl,"g_2"',sloped]\\
    & D
  \end{tikzcd}
  \caption{The biproduct diagram}
  \label{fig:biproduct}
\end{figure}

In a category with biproducts, that is, where the product and the coproduct coincide, we have the diagram from Figure~\ref{fig:biproduct}.
Continuing with $\mathbf{Vec}$ as a concrete example, we have $i_1(a) = (a,\vec 0)$, $i_2(b)=(\vec 0,b)$, and $[g_1,g_2](a,b)=g_1(a)+g_2(b)$.

In this category, the additive conjunction and the additive disjunction from
Linear Logic, with introduction rules given by:
\[
   \infer[\with_i]{\Gamma\vdash\pair tr:A\with B}{\Gamma\vdash t:A & \Gamma\vdash r:B}
   \qquad
   \infer[\oplus_i]{\Gamma\vdash\inl(t):A\oplus B}{\Gamma\vdash t:A}
   \qquad
   \infer[\oplus_i]{\Gamma\vdash\inr(t):A\oplus B}{\Gamma\vdash t:B}
\]
are both interpreted by the direct sum. The sum symbol is then interpreted
as the sum of vectors, and thus, $\inl(t)+\inr(r)$ is encoded as the
map $[i_1\circ t,i_2\circ r]\circ\Delta$.
\[
  \begin{tikzcd}[column sep=8mm,row sep=1mm]
    \hspace{2.1cm} &   \Gamma\ar[r,"\Delta"] & \Gamma\oplus\Gamma\ar[r,"{[i_1\circ t,i_2\circ r]}"] & A\oplus B\\
    & g\ar[r,mapsto] & (g,g)\ar[r,mapsto] & (t(g),\vec 0)+(\vec 0,r(g)) &[-9mm] = &[-9mm] (t(g),r(g))
  \end{tikzcd}
\]

Thus, $\inl(t)+\inr(r)$ is interpreted in the same way as $\pair tr$.
Indeed, $\pair tr$, as usual, is interpreted as $(t\oplus r)\circ\Delta$.
\[
  \begin{tikzcd}[column sep=8mm,row sep=1mm]
    \Gamma\ar[r,"\Delta"] & \Gamma\oplus\Gamma\ar[r,"t\oplus r"] & A\oplus B\\
    g\ar[r,mapsto] & (g,g)\ar[r,mapsto] & (t(g),r(g))
  \end{tikzcd}
\]

\subsection{Sup Without Sup}
The remark that $\inl(t)+\inr(r)$ is interpreted in the same way as $\pair tr$ was first made in~\cite{DiazcaroDowek25}, where we proposed to drop the $\odot$ connective by adding a new rule for disjunction introduction:
\[
  \infer[\vee_{i3}]{\Gamma\vdash \inlr tr:A\vee B}{\Gamma\vdash t:A & \Gamma\vdash r:B}
\]
This rule allows us to encode quantum computing without the need of a new connective. We showed in that draft how to make sense of this new rule in propositional logic, and in linear logic, how to encode quantum computing, but also how to use $\mathsf{inlr}$ as a general method to solve commuting cuts, when there is the introduction of a connective blocking a cut.

\subsubsection{The Non-Linear Case}
In another recent draft~\cite{DiazcaroMalherbe25} we consider the non-linear
case of the Sum and Scalar rules, and provided a categorical characterisation
for it.  Indeed, the categorical characterisation for the linear case from
Section~\ref{sec:catlin} is meaningful
only in the context of Linear Logic, where we can rely on categories such as
$\mathbf{Vec}$, which possess biproducts.  However, in the context of
propositional logic, the product and the coproduct are usually different, and
the interpretation of the sum of disjunctions cannot be the same as the
interpretation of the conjunction.  

In~\cite{DiazcaroMalherbe25}, we address this problem by proposing models for Sum and
Scalar rules within the framework of propositional logic, including
disjunctions, as well as all the other logical connectives. 

Let us focus first in the sum operator.
Since we were in the context of propositional logic, we started from the category
$\mathbf{Set}$, and sought for the simplest way to enrich its objects with an
operation so as to interpret the sum operator on it.  Thus, we first
considered the simpler algebraic structure, the category $\mathbf{Mag}$ of
magmas, whose objects are magmas and whose arrows are magma homomorphisms.
However, this category is not Cartesian closed, since the evaluation map is not
an homomorphism, and so it is not a good candidate.
Thus, we proposed studying the category $\TheCat$, whose
objects are magmas but whose arrows are functions from the $\mathbf{Set}$
category.
Naturally, it is Cartesian closed, since it is just the category
$\mathbf{Set}$ whose objects have an operation to which we do not require any
specific property. In addition to being Cartesian closed, not having
homomorphisms as arrows turns out to be necessary.
Indeed, without commutativity, and associativity, the sum operator is not
preserved by proof-terms, as shown in the following example.  Let $f$ be the
following proof-term.
\[
  \vdash{\lambda x.(\pi_1 x + \pi_2 x)}:(A\wedge A)\Rightarrow A
\]
Then, 
  $f\Big(\pair{{\color{red}a_1}}{{\color{red}a_2}}+\pair{{\color{blue}a_3}}{{\color{blue}a_4}}\Big)
  =f\pair{{\color{red}a_1}+{\color{blue}a_3}}{{\color{red}a_2}+{\color{blue}a_4}}
  =({\color{red}a_1}+{\color{blue}a_3})+({\color{red}a_2}+{\color{blue}a_4})$, while
  $f\pair{{\color{red}a_1}}{{\color{red}a_2}}+ f\pair{{\color{blue}a_3}}{{\color{blue}a_4}}=
  ({\color{red}a_1}+{\color{red}a_2})+({\color{blue}a_3}+{\color{blue}a_4})$.

The magmas provide the operation used to interpret the sum symbol. To our
surprise, disjunction in this calculus is not interpreted by a coproduct, as our
construction does not yield a coproduct unless the arrows in the category are
homomorphisms. Our proposal was to interpret disjunction \(A \vee B\) as the
union of the coproduct and the product in $\mathbf{Set}$, mimicking the idea of
a biproduct from the Linear Logic setting, that is, \(A \cp B = (A \uplus B)
\cup (A \times B)\).

When adding also scalars, 
we needed to consider a category \TheCatAlg of \emph{action magmas}, that is, magmas with an action map. The surprising aspect in both
cases, \TheCat and \TheCatAlg, is that we do not require algebraic properties in the categories beyond
equipping each object with an operation, or with an operation and an action
map. The disjunction is interpreted as \(A \cp B\), with its magma operator and
its action map performing most of the work, grouping the left side to the left
and the right side to the right, without requiring any associativity or
commutativity in the underlying sets. 

\subsection{Extensions and Future Directions}
\begin{figure}[t]
  \centering
  \begin{tikzcd}[column sep=3mm,row sep=0mm]
    \textrm{\color{red}\footnotesize Lambda-$S$}\ar[red,rd,out=0,in=90,dashed] & & & & \textrm{\color{red}\footnotesize $\mathcal{L^S_!}$}\ar[red,ld,out=180,in=90,dashed] \\
    & (\mathcal{S},\times,I)\ar[rr,bend left,"{(F,m)}"] & \bot & (\mathcal{V},\otimes,1)\ar[ll,bend left,"{(G,n)}"]
  \end{tikzcd}
  \caption{The duality between Lambda-$S$ and $\mathcal{L^S}$}
  \label{fig:duality}
\end{figure}
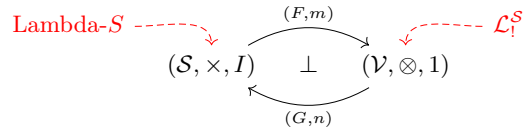

In~\cite{DiazcaroDowekIvniskyMalherbeWoLLIC2024}, we extended the $\mathcal{L^S}$ calculus to ILL by adding the modality ${!}$, resulting in the $\mathcal{L^S_!}$ calculus. Its model is currently under development for a forthcoming journal version of this paper. As expected, we employ the same adjunction as for Lambda-$S$ (cf.~Figure~\ref{fig:adjunction}), but interpret the calculus in a monoidal closed category and use the comonad $FG$ to interpret the ${!}$ modality (see Figure~\ref{fig:duality}).

In the same paper, we also added polymorphism to the calculus, and we are currently studying the interaction between polymorphism and the sums and scalar products in this setting, following the ideas of~\cite{Maneggia04}.

The formal connection between the Lambda-$S$ calculus and the $\mathcal{L^S_!}$ calculus remains an open research question. We are currently working on a formal proof of the equivalence between the two calculi and on the development of a realizability model for the $\mathcal{L^S}$ calculus, integrating ideas from~\cite{DiazcaroGuillermoMiquelValironLICS19}.

\section{Conclusion}
In this invited paper, we provided an overview of a research program with the
goal of extending the Curry-Howard-Lambek correspondence to quantum
computation.  We introduced the central ideas of two key frameworks. The
Lambda-$S$ calculus is a quantum lambda calculus that incorporates quantum
superpositions of lambda terms while enforcing linearity to address quantum
control. The $\mathcal{L^S}$ calculus, in turn, is a proof language
for intuitionistic linear logic, enhanced with novel rules to handle quantum
superpositions and scalar operations.

The Lambda-$S$ calculus incorporates quantum mechanical principles within the
lambda calculus framework, providing a structured approach to encoding quantum
programs.  In parallel, the $\mathcal{L^S}$ calculus extends
intuitionistic linear logic with scalar and sum operations, providing a logical
foundation for quantum computation and enabling categorical models that reveal
deep structural connections between logic and quantum computation.

These calculi bridge classical and quantum computational paradigms and suggest a fundamental duality between quantum computation and linear logic.
Future work includes further investigation of these foundational relationships, refinement and generalisation of the calculi, and the development of comprehensive categorical models.
Ultimately, these frameworks provide a promising formal foundation for quantum programming languages and offer deeper theoretical insight into quantum computation.

\subsubsection*{\ackname}
This work is supported by the European Union through the MSCA SE project QCOMICAL (Grant Agreement ID: 101182520), by the Plan France 2030 through the PEPR integrated project EPiQ (ANR-22-PETQ-0007), and by the Uruguayan CSIC grant 22520220100073UD.

\bibliographystyle{splncs04}
\bibliography{biblio}
\end{document}